\documentclass[10pt,conference]{IEEEtran}

\usepackage{cite}
\usepackage{epsfig}
\usepackage{graphicx}
\usepackage{amsfonts}
\usepackage{amsmath}
\usepackage{amssymb}
\usepackage{amsbsy}
\usepackage{bm}
\usepackage{multicol}
\usepackage{psfrag}
\usepackage{stfloats}
\usepackage{algorithmic}
\usepackage{algorithm}
\usepackage{array}
\usepackage{color}

\DeclareMathOperator{\km}{km}
\DeclareMathOperator{\dB}{dB}
\DeclareMathOperator{\dBm}{dBm}
\DeclareMathOperator{\Hz}{Hz}
\DeclareMathOperator{\MHz}{MHz}

\DeclareMathOperator{\Mbps}{Mbps}

\DeclareMathOperator*{\st}{s.t.}

\DeclareMathOperator*{\E}{\mathbb{E}}

\DeclareMathOperator{\maxo}{maximize}

\newcommand\qb{\ensuremath{\mathbf{q}}}
\newcommand\ssb{\ensuremath{\mathbf{s}}}

\newtheorem{Lemma}{Lemma}

\graphicspath{{figure/}}

\definecolor{orange}{RGB}{255,107,0}
\definecolor{green}{RGB}{0,160,20}

\begin{document}

\title{Robust Trajectory and Resource Allocation Design for Secure UAV-aided Communications}

\author{\IEEEauthorblockN{
		Xiaofang Sun\IEEEauthorrefmark{1}\IEEEauthorrefmark{2},
		Chao Shen\IEEEauthorrefmark{1}\IEEEauthorrefmark{2},
		Derrick Wing Kwan Ng\IEEEauthorrefmark{3}, and
		Zhangdui Zhong\IEEEauthorrefmark{1}\IEEEauthorrefmark{2}}
	
	\IEEEauthorblockA{
		\IEEEauthorrefmark{1}State Key Lab of Rail Traffic Control and Safety, Beijing Jiaotong University, Beijing, China\\
		\IEEEauthorrefmark{2}Beijing Engineering Research Center of High-speed Railway Broadband Mobile Communications, Beijing, China\\
		\IEEEauthorrefmark{3}School of Electrical Engineering and Telecommunications, University of New South Wales, Sydney, Australia}
	\IEEEauthorblockA{Email: \{xiaofangsun, chaoshen, zhdzhong\}@bjtu.edu.cn, w.k.ng@unsw.edu.au}\vspace{-5mm}}

\maketitle
\vspace{-5ex}

\vspace{-4ex}

\begin{abstract}
	This paper aims to enhance the physical layer security against potential internal eavesdroppings by exploiting the maneuverability of an unmanned aerial vehicle (UAV).
	We consider a scenario where two receivers with different security clearance levels require to be served by a legitimate transmitter with the aid of the UAV.
	We jointly design the trajectory and resource allocation to maximize the accumulated system confidential data rate.
	The design is formulated as a mixed-integer non-convex  optimization problem which takes into account the partial position information of a potential eavesdropper.
	To circumvent the problem  non-convexity, a series of transformations and approximations are proposed which facilitates the design of a computationally efficient suboptimal solution.
	Simulation results are presented to provide important system design insights  and demonstrate the advantages brought by the robust joint design for enhancing the physical layer security.
\end{abstract}

\vspace{-0.5ex}
\large\normalsize
\section{Introduction}

Recently, unmanned aerial vehicle (UAV) has drawn significant interests in wireless communications, due to its high mobility, on-demand of deployment, line-of-sight (LoS) air-to-ground channel, and low cost \cite{Yan_Solar_UAV_TCOM_2019}.
%It creates a fundamental paradigm shift in the fifth-generation (5G) and beyond 5G networks to enable fast and highly flexible deployment of communication infrastructures, especially in emergency scenarios.
%In particular, by exploiting the high maneuverability of the UAV, its trajectory can be flexibly designed to meet the heterogeneous quality-of-service (QoS) requirements.
Despite the special and unique advantages brought by UAVs, the open and strong LoS nature of air-to-ground channels lead to a high potential in information leakage.
As such, security plays an important role in UAV networks.
However, traditional encryption techniques require a large amount of energy consumption due to their high computational complexities \cite{PLS_Survey}.
As an alternative, physical layer security against potential eavesdroppings is an effective and energy efficient approach to address the security issues in UAV networks by exploiting the randomness of wireless channels from the information theoretic perspective.
Consequently, physical layer security has been widely studied in the literature for safeguarding wireless communications, e.g., \cite{MIMO_PLS_Survey}.
On the other hand, the high mobility and flexibility of UAVs can also be exploited to enhance the physical layer security.
Hence, a thorough study on the role of UAVs in secure communication systems is necessary.

Recently, the physical layer security in UAV systems has gained increasing attentions in the literature.
For example, UAV-aided jamming scheme was proposed to handle the security issues
in \cite{UAV_Jamming_CL_2019}.
Besides,  joint trajectory and resource allocation design was investigated in \cite{Xiaofang_GC,JointDesign_Access,Zhang_UAV_GC_2017} for improving the system secrecy rate with UAV-mounted base stations and UAV-aided relaying protocols, respectively.
However, the previous works are all based on an assumption that accurate position information of eavesdroppers is available at UAVs which is over optimistic.
In practice, obtaining such precise position information is challenging and expensive in terms of hardware cost and energy consumption.
Then, based on the imperfect channel state information of an external eavesdropper, a robust trajectory design was proposed  in \cite{Robust_UAV} for providing secure communications to the UAV-mounted legitimate transmitter.
Differently, in this work, we address the security issues caused by an internal eavesdropper with a lower security clearance level and  only partial position information available, via exploiting a UAV as a mobile relay.

Inspired by the advantages and challenges brought by UAVs, in this work, we study a \textit{robust} trajectory and resource allocation design for UAV-aided wireless communications to enhance the physical layer security against \textit{internal} eavesdroppings.
In the considered scenario, two ground receivers, named Bob and Eve, respectively, with different security clearance levels desire to be served by a remote transmitter (Alice) in the absence of direct communication links.
Specifically, the receiver with a lower security clearance level (Eve) not only imposes a certain QoS requirement on its own messages, but also intends to intercept the messages with a higher security clearance level.
To this end, a UAV is introduced as a mobile relay to meet the QoS requirements of different users and to guarantee secure communications simultaneously.
In many practical scenarios, obtaining accurate position information of a potential eavesdropper is usually difficult if not impossible.
As such, in this work, we exploit partial position information of the potential eavesdropper to facilitate the robust design by considering the worst-case.
The joint design of the trajectory and resource allocation is formulated a non-convex optimization problem to maximize the accumulated confidential data rate while ensuring Eve is only able to decode its desired data.
To facilitate the joint design, we propose a computationally efficient suboptimal iteration algorithm based on the successive convex approximation (SCA) approach \cite{Xiaofang_SDR}.
Moreover, we numerically examine the performance of the proposed robust joint design.

\textit{Notation:}
$\|\cdot\|$ denotes the Euclidean norm.
$[\mathbf{x}]_a$ denotes the $a$th component of vector $\mathbf{x}$.
The distribution of a circularly symmetric complex Gaussian (CSCG) variable with mean $\mu$ and variance $\sigma^2$ is denoted by $\mathcal{CN}(\mu,\sigma^2)$. $\mathbb{E}(\cdot)$ denotes the statistical expectation.
$(\cdot)^T$ denotes the transpose operation.

\vspace{-1ex}

\section{System Model and Problem Formulation}\label{sec:System}

\subsection{System Model}\label{Subsec:System_Model}

\begin{figure}[!t]
	\begin{center}
		\includegraphics[width=0.9\columnwidth]{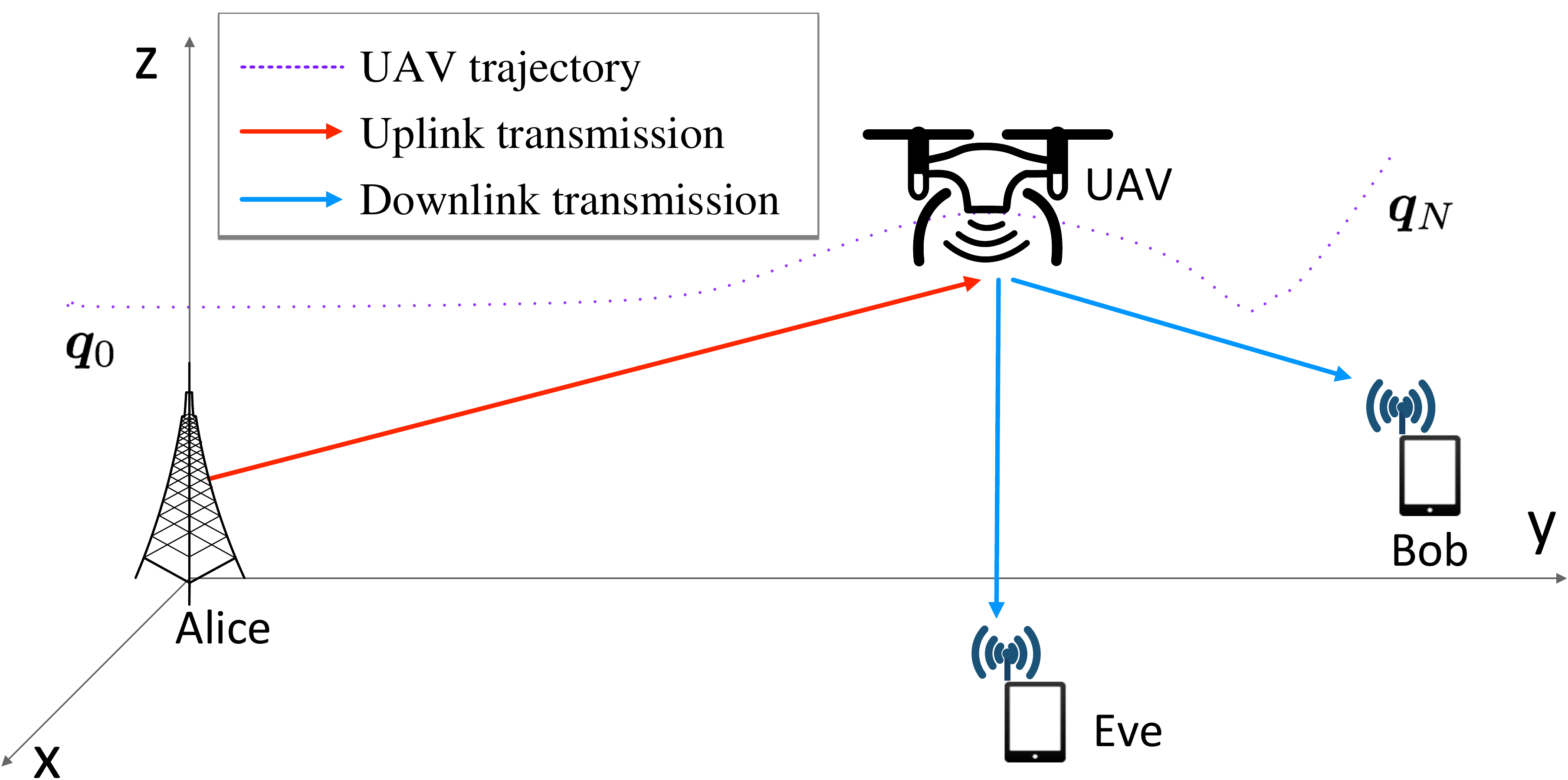}\vspace{-2ex}
		\caption{Illustration of a system model where Alice intends to serve Bob and Eve with the aid of UAV as a mobile relay.}\label{fig:system_model}
	\end{center}
\vspace{-5.5ex}
\end{figure}

In this paper, the considered secure UAV-aided communication system is modeled by a wiretap channel, as depicted in Fig. \ref{fig:system_model}, where Alice is the legitimate ground transmitter serving two ground users, named Bob and Eve, respectively, with different security clearance levels.
Without loss of generality, we assume that Bob has a higher security clearance level while Eve having a lower one.
In the considered scenario, the direct communication links from Alice to Bob and to Eve are both assumed to be not available.
To address this issue, a UAV is introduced to serve an intelligent relay providing the physical layer security against potential internal eavesdroppings via exploiting the UAV's maneuverability.
Moreover, we assume that the accurate locations of Alice and Bob are known by the UAV.
In contrast, only the partial position information of Eve is available for the trajectory and resource allocation design.

We define the maximum flight duration of the UAV by $T$, which mainly depends on the battery capacity, the flight control system, and the flight condition.
The three-dimensional (3D) location of the UAV at any certain time $t$ is denoted by $\qb(t)\in \mathbb{R}^3$.
Notably, the continuous UAV trajectory leads to infinite many variables, which make the problem intractable.
As a result, we adopt a commonly used approach in the literature to facilitate the system design.
In particular,  we discretize the maximum flight duration $T$ into $N$ time slots with equal time length $T_s$, i.e., $T=NT_s$.
Notably, each time slot is sufficiently small such that the flying speed of the UAV within it can be assumed to be constant.
Consequently, the UAV's coordinate at time slot $n$ can be denoted by $\qb_n\in \mathcal{R}^3$ for $n\in \{1,\cdots, N\}$.
The maximum speed of the UAV is denoted by $V_{\max}$.
Then, we have the maximum flight distance of each time slot, i.e., $D\triangleq T_sV_{\max}$.
For the sake of aviation safety, the initial and final locations of the UAV are generally predetermined, which are denoted by $\ssb_I$ and $\ssb_F$, respectively.
Besides, the height of  UAV is fixed at $H$.
Accordingly, we have the following constraints imposed on UAV in terms of trajectory
	\begin{align}\label{eq:Trajectory}
	\|\qb_{n}-\qb_{n-1}\|\leq D, \qb_0=\ssb_I,\qb_N=\ssb_F,[\qb_n]_3=H,~&\forall n.
	\end{align}

We assume that frequency division duplex (FDD) protocol is adopted.
Specifically,  the uplink (UL) transmission (from Alice to UAV) and the downlink (DL) transmission (from UAV to Bob and Eve) are separated in two orthogonal frequency bands.
In addition, the communication bandwidth can be dynamically optimized and allocated to improve the spectral efficiency.
Denote the total bandwidth by $B\Hz$, which is adaptively divided into $b_{1,n}\Hz$ and $b_{2,n}\Hz$ for uplink and downlink transmissions, respectively, in time slot $n$. Hence, we have
\begin{align}\label{eq:st_BandwidthAllocation}
b_{1,n}\ge 0,~b_{2,n}\ge 0,~b_{1,n}+b_{2,n}\leq B,~\forall n.
\end{align}
The UAV adopts the cache-aided decode-and-forward (DF) protocol for information relaying.
Here, we assume that the cache equipped at the UAV is large enough such that there will be no packet overflow.

Denote the locations of Alice and Bob in a 3D Cartesian coordinate system by $\ssb_{a}$ and $\ssb_{b}$, respectively.
Without loss of generality, the uncertain ground area of Eve is modeled by a disc with radius $d_{e}$ centered at $\ssb_{e}$.
According to the filed trial results in \cite{3GPP}, the line-of-sight (LoS) component dominates air-to-ground channels in many practical scenarios, especially for rural areas or at moderately high UAV altitude.
Hence, the air-to-ground channel between UAV and a ground node $k$ at each time slot can be accordingly modeled by
\begin{align}\label{eq:ChannelGain}
h_{k,n}=\frac{\gamma_0}{\|\qb_n-\ssb_{k}\|^2}, ~~ k \in \{a,b\},
\end{align}
%which are assumed to be available at the UAV,
where $\gamma_0$ is the channel power gain at a reference distance of $1$ m.
To guarantee the communication security, we focus on the worst-case scenario.
In particular, the channel gain between the UAV and Eve is bounded by
\begin{align}\label{eq:ChannelGain_Eve}
	h_{e,n}^s = \frac{\gamma_0}{d_{e,n}^s} ~\text{and}~
	h_{e,n}^w = \frac{\gamma_0}{d_{e,n}^w},
\end{align}
which are the upper bound and lower bound, respectively, where
\begin{align}\label{eq:Distance_Eve}
d_{e,n}^s&=\|\qb_{n}-\ssb_{e}\|^2+d_{e}^2-2d_{e}D_{G,n},\\
d_{e,n}^w&=\|\qb_{n}-\ssb_{e}\|^2+d_{e}^2+2d_{e}D_{G,n},
\end{align}
where $D_{G,n}=\sqrt{([\qb_{n}]_1-[\ssb_{e}]_1)^2+([\qb_{n}]_2-[\ssb_{e}]_2)^2}$ is the projected distance on the ground.

\subsection{Transmission Strategy}

Denote the uplink signal in time slot $n$ from Alice to the UAV by $x_{a,n}$£¬ which contains the messages intended to Bob and/or Eve. The average transmission power satisfies
\begin{align}\label{eq:st_power_alice}
p_{a,n}\triangleq \E[|x_{a,n}|^2] \leq P_{\mathrm{A}}, ~~n\in \{1,\cdots, N\}.
\end{align}

The received signal at the UAV in time slot $n$ is given by
\begin{align}\label{eq:y_u}
y_{u,n}=\sqrt{h_{a,n}}x_{a,n}+z_{u,n},
\end{align}
where $z_{u,n}\sim \mathcal{CN}(0,\sigma^2)$ denotes the additive white Gaussian noise (AWGN).
Without loss of generality, we assume that the received noise powers at all nodes are the same.
 % at all nodes and each time slot
Then, the uplink capacity from Alice to the UAV in time slot $n$ is
\begin{align}\label{eq:Rate_uav}
C^{\mathrm{A}\rightarrow \mathrm{U}}_n= b_{1,n}\ln\left(1+\frac{\gamma p_{a,n}}{b_{1,n}\|\qb_{n}-\ssb_{a}\|^2}\right)\text{nat/s},
\end{align}
where $\gamma = \gamma_0/\sigma^2$ is the noise normalized channel power gain.

We denote  $R_{u,n}^b$ and $R_{u,n}^e$ as the coding rates from Alice to the UAV for conveying messages to  Bob and Eve, respectively. Thus, $R_{u,n}^b$ and $R_{u,n}^e$ are constrained by the achievable uplink capacity, i.e.,
\begin{align}\label{eq:st_UplinkRate}
R_{u,n}^b\geq 0,~ R_{u,n}^e\geq 0,~R_{u,n}^b+R_{u,n}^e\leq C^{\mathrm{A}\rightarrow \mathrm{U}}_n, ~\forall n.
\end{align}

At the UAV, the received data from Alice is cached in the buffer for forwarding to Bob and Eve separately in the following time slots.
Recall that Bob and Eve are both desired to be served but with different secrecy clearance levels.
To this end, the two types of signals are transmitted in orthogonal time slots.
For the sake of clarity, we introduce a set of binary variables $\rho_n$, where
\begin{align}\notag
\rho_n\triangleq\begin{cases}
1, &\text{the UAV only serves Bob},\\
0, &\text{the UAV only serves Eve}.
\end{cases}
\end{align}

\subsubsection{$\rho_n=1$}
The UAV transmits confidential signal $x_{b,n}$ to Bob but has a potential of leakage to Eve.
The transmission power is defined by $p_{b,n}\triangleq \E[|x_{b,n}|^2]$.
The coding rate of $x_{b,n}$ is defined by $R_{b,n}~\text{nat/s}$.
Due to the open nature of wireless channels, the transmitted signal can be received at both Bob and Eve. Their received signal equations are given by $y_{k,n}=\sqrt{h_{k,n}}x_{b,n}+z_{k,n}$, where $k\in \{b,e\}$. The corresponding channel capacity between the UAV and  Bob is
\begin{align}\label{eq:Rate_Bob}
C^{\mathrm{U}\rightarrow \mathrm{B}}_{n}= \rho_n b_{2,n}\ln\left(1+\frac{\gamma p_{b,n}}{b_{2,n}\|\qb_{n}-\ssb_{b}\|^2}\right)\text{nat/s}.
\end{align}
The potential maximum wiretap channel capacity between the UAV and Eve is given by
\begin{align}\label{eq:Rate_Bob_at_Eve}
C^{\mathrm{U}\rightarrow \mathrm{E}}_{b,n}= \rho_nb_{2,n}\ln\left(1+\frac{\gamma p_{b,n}}{b_{2,n}d_{e,n}^s}\right)\text{nat/s}.
\end{align}

To ensure the communication reliability at Bob, the coding rate $R_{b,n}$ is constrained by $R_{b,n}\leq C^{\mathrm{U}\rightarrow \mathrm{B}}_{n}$. Moreover, the correspondingly minimum secrecy rate for $x_{b,n}$ in time slot $n$ is defined as
\begin{align}\label{eq:SecrecyRate}
R_{n}^{s}\triangleq [R_{b,n}-C^{\mathrm{U}\rightarrow \mathrm{E}}_{b,n}]^+.
\end{align}

\subsubsection{$\rho_n=0$}
The signal $x_{e,n}$ is intended to Eve from UAV in time slot $n$ with $\rho_n=0$.
Its allocated power and coding rate are defined by $p_{e,n}\triangleq \E[|x_{e,n}|^2]$ and $R_{e,n}~\text{nat/s}$, respectively.
Likewise the received signal with $\rho_n=1$, for guaranteeing reliable communications, we consider the potential minimum channel capacity of $x_{e,n}$ at Eve,  which is given by
\begin{align}\label{eq:Rate_Eve}
C^{\mathrm{U}\rightarrow \mathrm{E}}_n= (1-\rho_n)b_{2,n}\ln\left(1+\frac{\gamma p_{e,n}}{b_{2,n}d_{e,n}^w}\right)\text{nat/s}.
\end{align}
Similarly, to ensure communication reliability at Eve, the coding rate $R_{e,n}$ is constrained by $R_{e,n}\leq C^{\mathrm{U}\rightarrow \mathrm{E}}_n$.

Recall that the UAV acts as an aerial relay to cache and forward data from Alice to Bob and Eve. Hence, additional information causality constraints \cite{JointDesign_Access} in terms of signals to Bob and Eve are imposed on UAV, which are given by\vspace{-1ex}
	\begin{align}\label{eq:st_InformationCausality}
	\sum_{i=1}^n R_{b,i} \leq \sum_{i=1}^n R_{u,i}^b ~\text{and}~
	\sum_{i=1}^n R_{e,i} \leq \sum_{i=1}^n R_{u,i}^e, \forall n,
	\end{align}
respectively.

We note that the allocated transmission powers $p_{b,n}$ and $p_{e,n}$ are limited by the maximum transmission powers of UAV, i.e.,
\begin{align}\label{eq:UAV_Peak_Power}
&0\leq p_{b,n} \leq P_{\mathrm{U}}, ~0\leq p_{e,n} \leq P_{\mathrm{U}},~\forall n.
\end{align}

%In addition, we also consider the long-term average power budget imposed on UAV, which can be characterized by\vspace{-0.5ex}
%\begin{align}\label{eq:st_power_UAV}
%\sum_{n=1}^N \left(p_{b,n}+p_{e,n}\right)&\leq N \overline{P}_U,
%\end{align}\vspace{-0.5ex}
%where $\overline{P}_U$ is the average output power limit of the UAV.

\subsection{Problem Formulation}
In this section, we formulate the resource allocation and trajectory design as an optimization problem.
In the considered system, the UAV assists Alice to deliver a certain amount of data to Eve for a given time duration $T$.
We aim to maximize the accumulated confidential data rate achieved at Bob while guaranteeing a minimum required data rate for Eve to decode its own messages.
To this end, the resource allocation and trajectory of UAV are to be jointly optimized.
Mathematically, the optimization problem is formulated as\vspace{-1ex}
\begin{subequations}\label{eq:Opt_OMA}
	\begin{align}
	\texttt{P1:}~~\underset{\Delta}{\maxo}~~&\sum_{n=1}^N R_n^s \label{eq:Obj_SecData}\\
	\st~~
	&C^{\mathrm{U}\rightarrow \mathrm{B}}_{n}\geq R_{b,n},\forall n,\label{eq:st_OMA_Rate_Bob}\\
	&C^{\mathrm{U}\rightarrow \mathrm{E}}_{n}\geq R_{e,n}, \forall n,\label{eq:st_OMA_Rate_Eve}\\
	&\sum_{n=1}^NR_{e,n} \geq R_{\mathrm{E}},\label{eq:st_Accumulated_Eve_Rate}\\
	&\rho_n \in \{0,1\}, \forall n,\label{eq:st_Binary_Constraint}\\
	&\eqref{eq:Trajectory},\eqref{eq:st_BandwidthAllocation},\eqref{eq:st_power_alice},\eqref{eq:st_UplinkRate},\eqref{eq:st_InformationCausality},\eqref{eq:UAV_Peak_Power},\notag
	\end{align}
\end{subequations}
where $\Delta\triangleq \{\Delta_1,\{b_{2,n},p_{a,n},p_{b,n},p_{e,n},\rho_n\}_{n=1}^N \}$ and $\Delta_1\triangleq \{b_{1,n},\qb_{n},R_{u,n}^b,R_{u,n}^e,
R_{b,n},R_{e,n}\}_{n=1}^N$ are the sets of optimization variables.
$R_{\mathrm{E}}$ is the minimum required data rate for Eve to decode its own message.
The considered optimization formulation is a generalization of existing ones in the literature.
For example, external eavesdropping, e.g., \cite{Zhang_UAV_GC_2017}, and complete position information of the eavesdropper, e.g., \cite{Xiaofang_GC}, are two special cases of this framework with $R_{\mathrm{E}}=0$ and $d_{e}=0$, respectively.

\section{Joint Trajectory and Resource Allocation Design Strategy}\label{Sec:JointDesign}

The considered optimization problem \texttt{P1} is a mixed-integer programming which is NP-hard in general \cite{Derrick_TWC_2016}.
In particular, the optimization variables are coupled together resulting in an intractable problem.
To tackle \texttt{P1} in \eqref{eq:Opt_OMA}, we first provide some useful insights to simplify the problem at hand.
Then, we propose a series of transformations and approximations to facilitate the design of a suboptimal solution.

\subsection{Simplifications of the Optimization Problem \texttt{P1}}

In particular, without loss of optimality, the uplink and downlink transmissions in each time slot fully occupy the total bandwidth, i.e., $b_{1,n}+b_{2,n}=B,\forall n$. This is due to the fact that if the bandwidth allocation satisfies  $b_{1,n}+b_{2,n} < B$ with strict inequality  in some time slots, the objective function may be further increased by allocating the available bandwidth to the uplink and/or downlink channels subject to the constraints.

Then, to further facilitate the joint design, in the following lemma, we discuss the optimal transmission strategy for secure communications adopted at the UAV.
\begin{Lemma}
	If the UAV locates closer to Bob than Eve, i.e., $\|\qb_{n}-\ssb_{b}\|^2< \|\qb_{n}-\ssb_{e}\|^2+d_{e}^2-2d_{e}D_{G,n}$, the optimal strategy is to  transmit confidential signal with $\rho_n=1$. Otherwise, the UAV either caches data from the uplink channel or delivers data with a lower security clearance level to Eve.
	\begin{IEEEproof}
		This lemma can be proved by contradiction.
		Suppose that the UAV transmits the confidential signal when it is closer to Eve than Bob, i.e., $\|\qb_{n}-\ssb_{b}\|^2\geq  \|\qb_{n}-\ssb_{e}\|^2+d_{e}^2-2d_{e}D_{G,n}$, in certain time slots and achieves the optimal solution.
		Based on the definition of the secrecy rate in \eqref{eq:SecrecyRate}, the objective value keeps invariant even without power and bandwidth allocation in these time slots.
		Hence, the original resources dedicated to these time slots can be saved and be used in other time slots to improve the system performance.
	\end{IEEEproof}
\end{Lemma}

By exploiting this lemma, the objective function  given in \eqref{eq:Obj_SecData} can be simplified to
\vspace{-2ex}
\begin{align}\label{eq:st_AccumulatedSecData_Simplied}
\sum_{n=1}^N R_{b,n}-C^{\mathrm{U}\rightarrow \mathrm{E}}_{b,n}.
\end{align}

We note that $\rho_n$ is a binary optimization variable in \texttt{P1} and additionally coupled with other variables.
Now, by exploiting the binary property of $\rho_n$, $C^{\mathrm{U}\rightarrow \mathrm{B}}_n$ in  \texttt{P1} can be equivalently written as:
\begin{align}
C^{\mathrm{U}\rightarrow \mathrm{B}}_n=\rho_nb_{2,n}\ln\left(1+\frac{\gamma p_{b,n}}{\rho_nb_{2,n}\|\qb_n-\ssb_{b}\|^2}\right).
\end{align}
This can be proved directly when $\rho_n=1$ and via the L'Hospital's rule when $\rho_n=0$.

Furthermore, we introduce two auxiliary optimization variables $\eta_n\geq 0$ and $\tau_n\geq 0$ to replace $\rho_nb_{2,n}$ and $(1-\rho_n)b_{2,n}$, respectively.
To this end, two types of constraints are imposed on the new variables, which are given by
\begin{subequations}\label{eq:st_eta_tau}
	\begin{align}
\tau_n=B-b_{1,n}-\eta_n, ~\rho_n\geq 0,~\eta_n\geq 0, ~\forall n,\label{eq:st_eta_tau_sum}\\
\eta_n\tau_n\leq 0,~ \forall n.\label{eq:st_eta_tau_times}
\end{align}
\end{subequations}
These constraints guarantee that the UAV serves Bob and Eve with orthogonal time slots.
Notably, $\eta_n$ and $\tau_n$ are coupled in \eqref{eq:st_eta_tau_times} which is non-convex and to be addressed in the following sections.

\subsection{Transformation of the Optimization Problem \texttt{P1}}
We note that the optimization variables are still coupled in constraints \eqref{eq:st_UplinkRate}, \eqref{eq:st_OMA_Rate_Bob}, \eqref{eq:st_OMA_Rate_Eve}, \eqref{eq:st_AccumulatedSecData_Simplied}, \eqref{eq:st_eta_tau_times}.
In this section, we transform the optimization problem \texttt{P1} into its equivalent form to facilitate the computationally efficient resource allocation and trajectory design.
To this end, we first introduce some auxiliary optimization variables $\mu_{a,n}$, $\mu_{b,n}$, and $\mu_{e,n}$ to bound $\|\qb_{n}-\ssb_{a}\|^{2}$, $\|\qb_{n}-\ssb_{b}\|^{2}$, $d_{e,n}^w$ from above, respectively.
As a consequence, some additional constraints are imposed on the auxiliary variables, which are given by
\begin{subequations}\label{eq:st_mu}
	\begin{align}
	\|\qb_{n}-\ssb_{a}\|^{2} \leq \mu_{a,n}, ~\mu_{a,n}\geq 0,~&\forall n,\\
	\|\qb_{n}-\ssb_{b}\|^{2} \leq \mu_{b,n},~\mu_{b,n}\geq 0, ~&\forall n,\\
	d_{e,n}^w \leq \mu_{e,n},~\mu_{e,n}\geq 0, ~&\forall n.
	\end{align}
\end{subequations}

Inspired by the fact that a quadratic-over-linear function is convex  with respect to its positive inputs, respectively \cite{Boyd_2004}, we introduce some new variables $\alpha_{a,n}\geq 0$, $\alpha_{b,n}\geq 0$, and $\alpha_{e,n}\geq 0$ to replace $\sqrt{p_{a,n}}$, $\sqrt{p_{b,n}}$, and $\sqrt{p_{e,n}}$, respectively.

We then introduce a set of auxiliary variables $\theta_n\geq 0$ to bound $\frac{\alpha_{b,n}^2}{d_{e,n}^s}$ from above.
As such, the constraint imposed on the new variable is given by
\begin{align}\label{eq:st_theta}
\frac{\alpha_{a,n}^2}{\theta_n}+2d_{e}D_{G,n}\!\leq\! \|\qb_{n}-\ssb_{e}\|^2+d_{e}^2,~\theta_n\geq 0,~\forall n.
\end{align}

Based on the previous transformations, the optimization problem \texttt{P1} is transformed into its equivalent form, which is given by
\begin{subequations}\label{eq:Opt_Transformed}
	\begin{align}
\texttt{P2:}	\underset{\Delta_2}{\maxo}~&\sum_{n=1}^N R_{b,n}-\eta_n\ln\left(1+\frac{\gamma \theta_n}{\eta_n}\right)\label{eq:Obj}\\
	\st~
	&R_{u,n}^b\!\!+\!\!R_{u,n}^e \!\leq\! b_{1,n}\ln\left(\!\!1+\frac{\gamma \alpha_{a,n}^2}{b_{1,n}\mu_{a,n}}\!\!\right), \forall n, \label{eq:st_IC}\\
	&\eta_n\ln\left(1+\frac{\gamma \alpha_{b,n}^2}{\eta_n \mu_{b,n}}\right)\geq R_{b,n}, \forall n,\label{eq:st_RC_Bob}\\
	&\tau_n \ln\left(1+\frac{\gamma \alpha_{e,n}^2}{\tau_n \mu_{e,n}}\right)\geq R_{e,n}, \forall n, \label{eq:st_RC_Eve}\\
	&\alpha_{a,n}^2\leq P_{\mathrm{A}}, ~\alpha_{b,n}^2\leq P_{\mathrm{U}}, ~\alpha_{e,n}^2\leq P_{\mathrm{U}},~ \forall n,\label{eq:st_Power}\\
	&
	\eqref{eq:Trajectory},
	\eqref{eq:st_BandwidthAllocation},
	\eqref{eq:st_InformationCausality},
	\eqref{eq:st_Accumulated_Eve_Rate},
	\eqref{eq:st_eta_tau},
	\eqref{eq:st_mu},
	\eqref{eq:st_theta},
	\notag
	\end{align}
\end{subequations}
where $\Delta_2\!\!\triangleq\!\!\{\Delta_1\!,\!\{\!\alpha_{a,n},\!\alpha_{b,n},\!\alpha_{e,n},\!\tau_n,\!\eta_n,\!\mu_{a,n},\!\mu_{b,n},\!\mu_{e,n},\!\theta_n\!\}_{n=1}^N\!\}$ is the new optimization variable set.
However, the objective function \eqref{eq:Obj} and constraints \eqref{eq:st_eta_tau_times}, \eqref{eq:st_theta}, \eqref{eq:st_IC}, \eqref{eq:st_RC_Bob}, and \eqref{eq:st_RC_Eve}  are still non-convex, due to the coupling variables.
In the following section, we focus on the approximations of these non-convex objective function and constraints.

\subsection{Approximations of the Optimization Problem \texttt{P2}}

In this section, we convexify the non-convex objective function and constraints via the SCA approach \cite{Note_SCA}.

\subsubsection{SCA Method Based on the First-Order Taylor Expansion}

We note that the non-convex functions in the constraints  \eqref{eq:st_IC}, \eqref{eq:st_RC_Bob}, and \eqref{eq:st_RC_Eve} are desired to be approximated to concave functions. To this end, the SCA approach based on the first-order Taylor expansion is employed.
In particular, $\frac{\alpha_{k,n}^2}{\mu_{k,n}}$ is convex with respect to $\alpha_{k,n}$ and $\mu_{k,n}$, respectively, where $k\in \{a,b,e\}$.
Thus, for any fixed point in the $r$th iteration $\alpha_{k,n}^r\geq 0$ and $\mu_{k,n}^r\geq 0$, $\frac{\alpha_{k,n}^2}{\mu_{k,n}}$ can be bounded from below, i.e., $\frac{\alpha_{k,n}^2}{\mu_{k,n}}\geq \tilde{R}_{k,n}\triangleq \frac{2\alpha_{k,n}^r}{\mu_{k,n}^r}\alpha_{k,n}-\frac{(\alpha^r_{k,n})^2}{(\mu_{k,n}^r)^2}\mu_{k,n}$.
As a result, the constraints  \eqref{eq:st_IC}, \eqref{eq:st_RC_Bob}, and \eqref{eq:st_RC_Eve} are approximated to
\begin{subequations}\label{eq:st_Rate_Convex}
	\begin{align}
	b_{1,n}\ln\left(1+\frac{\gamma \tilde{R}_{a,n}}{b_{1,n}}\right)\geq R_{u,n}^b+R_{u,n}^e,&\forall n,\label{eq:st_Rate_Alice_Convex}\\
	\eta_n\ln\left(1+\frac{\gamma \tilde{R}_{b,n}}{\eta_n}\right)\geq R_{b,n},&\forall n,\label{eq:st_Rate_Bob_Convex}\\
	\tau_n\ln\left(1+\frac{\gamma \tilde{R}_{e,n}}{\tau_n}\right)\geq R_{e,n},&\forall n,\label{eq:st_Rate_Eve_Convex}
	\end{align}
\end{subequations}
respectively, which are convex, since perspective functions preserve convexity \cite{Boyd_2004}.

Moreover, the second term on the left-hand side of the objective function \eqref{eq:Obj} desires to be approximated to a convex function and the associated constraint \eqref{eq:st_theta} is to be approximated to a concave one.
To this end, by employing the first-order Taylor expansion, the objective function \eqref{eq:Obj} and the constraint are approximated to
	\begin{align}
	&\overline{R}_n^s\triangleq R_{b,n}-\eta_n\ln\left(1+\frac{\gamma \theta_n^r}{\eta_n^r}\right)-\gamma \frac{-\frac{\theta_n^r\eta_n}{\eta_n^r}+\theta_n}{1+\gamma \frac{\theta_n^r}{\eta_n^r}},\notag\\
	&\frac{\alpha_{b,n}^2}{\theta_n}\!+\!2d_{e}D_{G,n}-d_{e}^2\!\leq\! \|\qb_{n}^r\!-\!\ssb_{e}\!\|^2\!+\!2\mathbf{r}(\qb_{n}\!-\!\qb_{n}^r),\forall n.\label{eq:st_theta_Convex}
	\end{align}
where $\mathbf{r}\triangleq (\qb_{n}^r)^T-\ssb_{e}^T$. $\theta_n^r$,  $\eta_n^r$, and $\qb_{n}^r$ are fixed points in the $r$th iteration.

\subsubsection{SCA Method Based on Arithmetic-Geometric Mean (AGM) Inequality}

Notably, constraint \eqref{eq:st_eta_tau_times}  is bilinear with respect to $\tau_n$ and $\eta_n$, respectively.
For handling the bilinear function, we adopt the SCA approach based on AGM inequality.
As such, the non-convex bilinear function is sequentially upper bounded by a convex one, i.e.,
$\eta_n\tau_n\!\!\leq\!\! \frac{1}{2}\left(\left(\frac{\eta_n}{\psi_n}\right)^2\!+\!\left(\tau_n\psi_n\right)^2\right), \forall n$,
where $\psi_n\geq 0$ is a fixed point for tightening the upper bound, which is updated by $\psi_n^{r+1}=\sqrt{\frac{\eta_n^r}{\tau_n^r}}$ in the $r$th iteration.
Consequently, the non-convex constraint \eqref{eq:st_eta_tau_times} can be safely replaced by
\begin{align}\label{eq:st_eta_tau_times_convex}
\frac{1}{2}\left(\left(\frac{\eta_n}{\psi_n}\right)^2+\left(\tau_n\psi_n\right)^2\right)\leq 0, \forall n.
\end{align}
Specifically, we find that the non-convex feasible set is approximated by a convex set which only contains one possibility, i.e., $\eta_n=0$ and $\tau_n=0,\forall n$.
However, this possible solution violates constraint \eqref{eq:st_Accumulated_Eve_Rate} for guaranteeing the QoS requirement of Eve.
To address this issue, we augment \eqref{eq:st_eta_tau_times_convex} into the objective function via introducing a set of penalty factor $\lambda_n\gg 0$ to obtain an equivalent form.

Based on the previous approximations, the optimization problem \texttt{P2} in \eqref{eq:Opt_Transformed} has can be approximated to a convex one, which is given by\vspace{-1ex}
\begin{subequations}\label{eq:Opt_Convex}
	\begin{align}
	\texttt{P3:}~\underset{{\Delta_2}}{\maxo}~&R_S\triangleq \sum_{n=1}^N \overline{R}_n^s\hspace*{-0.5mm}-\hspace*{-0.5mm}\sum_{n=1}^N\lambda_n\left(\left(\frac{\eta_n}{\psi_n}\right)^2\hspace*{-0.5mm}+\hspace*{-0.5mm}\left(\tau_n\psi_n\right)^2\right)\notag\\
	\hspace*{-0.5mm}\st~	&\hspace*{-0.5mm}
	\eqref{eq:Trajectory},\hspace*{-0.075mm}
	\eqref{eq:st_BandwidthAllocation}\hspace*{-0.075mm},
	\eqref{eq:st_InformationCausality}\hspace*{-0.075mm},
	\eqref{eq:st_Accumulated_Eve_Rate}\hspace*{-0.075mm},
	\eqref{eq:st_eta_tau_sum}\hspace*{-0.05mm},
	\eqref{eq:st_mu}\hspace*{-0.05mm},
	\eqref{eq:st_Power}\hspace*{-0.05mm},
	\eqref{eq:st_Rate_Convex}\hspace*{-0.05mm},
	\eqref{eq:st_theta_Convex}. \notag
	\end{align}
\end{subequations}

The above optimization problem \texttt{P3} is convex given the fixed points and can be solved efficiently by off-the-shelf convex solvers, e.g., CVX \cite{cvx}.

We note that $\lambda_n$ is updated by adopting gradient descent method \cite{Note_Subgradient}. Specifically, at the $r$-th iteration, $\lambda_n^r$ is updated by
$
\lambda_n^{r+1} = \lambda_n^r+\delta_n\left(\left(\frac{\eta_n^r}{\psi_n^r}\right)^2+\left(\tau_n^r\psi_n^r\right)^2\right)
$,
where $\delta_n\geq 0$ is the $r$th step size.

Based on the concept of SCA \cite{Xiaofang_SDR}, the fixed points are iteratively updated in the $r$th iteration as
\begin{subequations}\label{eq:Update}
	\begin{align}
	&\alpha_{k,n}^r=\alpha_{k,n},\mu_{k,n}^{r}=\mu_{k,n},~\forall n, k\in\{a,b,e\},\notag\\
	&\theta_n^{r}=\theta_n,~\eta_n^r=\eta_n,~ \qb_{n}^r=\qb_{n}, ~\forall n.\notag
	\end{align}
\end{subequations}
Therefore, we obtain the SCA-based iterative algorithm for problem \eqref{eq:Opt_Convex}, as summarized in Algorithm \ref{Algorithm_SCA}, in which we define $\Gamma_1^r$ as
\begin{align}\notag
\Gamma_1^r\triangleq \{\{\alpha_{a,n}^r,\alpha_{b,n}^r,\alpha_{e,n}^r,\mu_{a,n}^r,\mu_{b,n}^r,\mu_{e,n}^r,\theta_n^r,\eta_n^r,\qb_n^r\}_{n=1}^N \}.
\end{align}

Based on the results in  \cite{BSUM}, the proposed SCA-based algorithm can converge to a stationary Karush-Kuhn-Tucker point.
Hence, the algorithm is able to achieve a suboptimal solution of \texttt{P1} with polynomial-time computational complexity.

 \begin{algorithm}[!tb]
	\caption{SCA-based Algorithm for Joint Trajectory and Resource Allocation Design}
	\label{Algorithm_SCA}
	\begin{algorithmic}[1]\small
		\STATE {\bf Initialize} $\Gamma_1^0$, $\epsilon=1$,
		and iteration index $r=1$.
		\WHILE {$\epsilon\geq 10^{-4}$}
		\STATE Update  $\Delta_2$ with fixed $\Gamma_1^r$ by \eqref{eq:Opt_Convex};\\
		\STATE Update $\Gamma_1^{r+1}$ based on $\Delta_2$;\\
		\STATE Update $\epsilon=\left|R_{S}^{(r)}-R_{S}^{(r-1)}\right|/R_{S}^{(r-1)}$;
		\STATE Update $r=r+1$;
		\ENDWHILE
		\STATE {\bf Output} $\Delta_3$ and $R_S$
	\end{algorithmic}
\end{algorithm}

\section{Numerical Results}\label{sec:simulation}
In this section, we numerically examine the secure performance of the proposed transmission strategy.
In the simulation, all nodes are located in a 3D Cartesian coordinate system.
Without loss of generality, Alice and Bob are set to be located at (0, 0, 0) and (5 km, 0, 0), respectively.
The initial and final locations of the UAV are located at (-2 km, 1 km, 0.1 km) and (6 km, 1 km, 0.1 km), respectively.
The reference channel  power gain at distance one meter and the noise power
are set as  $\gamma_0=-50~\dB$ and $\sigma^2=-150~\dBm/\Hz$, respectively.
The communication bandwidth is set as $B= 10~\MHz$.
The maximum flight duration is set as $T=450~\text{s}$.
The maximum speed of the UAV and the time slot duration are set as $V_{\max}=20~\text{m/s}$ and $T_s=10~s$, respectively.
The maximum power budget at Alice and the UAV are  set as $P_{\mathrm{A}}=30~\dBm$ and $P_{\mathrm{U}}=27~\dBm$, respectively.

\begin{figure}[!t]
	\begin{center}\vspace{-1ex}
		\includegraphics[width=3.5 in]{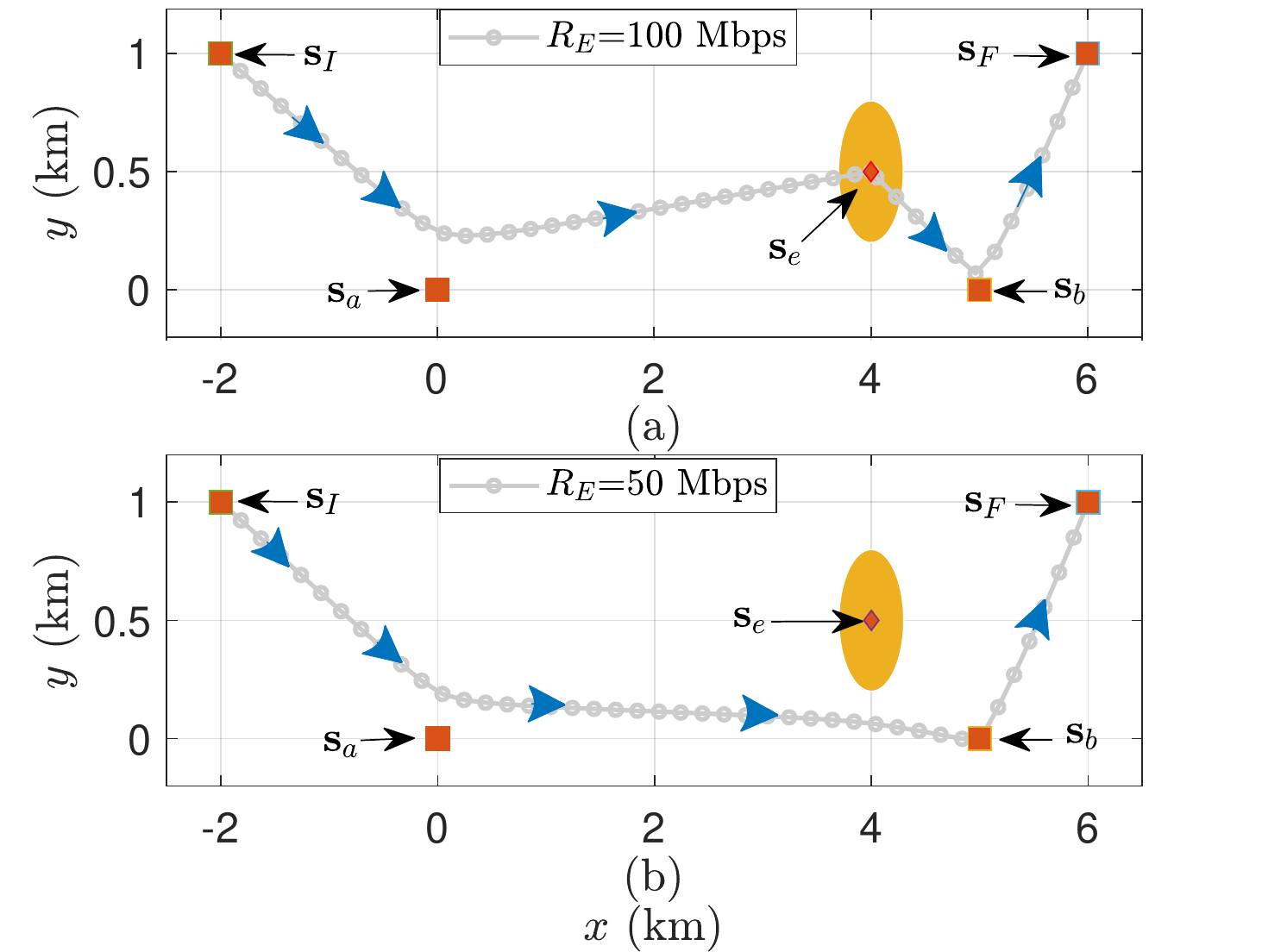}\vspace{-1ex}
		\caption{Trajectory of the UAV with different targets data rate of Eve, i.e., $R_{\mathrm{E}}= 100~\Mbps$ and $R_{\mathrm{E}}=50~\Mbps$ with $\ssb_{e}=(4~\km, 0.5~\km, 0)$, $d_{e}=0.3~\km$.}\label{fig:Trajectory}
	\end{center}
\vspace{-4ex}
\end{figure}

In Fig. \ref{fig:Trajectory}, we plot the optimized trajectory of the UAV against the potential internal eavesdropping.
For the sake of observation, the projected ground locations of Alice, Bob, initial and final points of UAV are displayed with makers while the uncertain area of Eve is shaded.
From this figure, we can observe that the UAV first tries to approach Alice for data caching as much as possible, and then flies towards Eve for completing data delivery mission.
After that, UAV flies away Eve to Bob and hovers above Bob for secure communications with the best-effort.
Finally, UAV flies back to its final location with its full speed.
Moreover, we find that when the accumulated data rate requirement of Eve becomes more stringent, the UAV has to spend more time and power resources on serving Eve.
For example, the UAV needs to hover above the central point of Eve's uncertain area when the data requirement of Eve is high, e.g., $R_{\mathrm{E}}=100~\Mbps$, while only flying close to Eve's uncertain area with a relatively high speed when $R_{\mathrm{E}}=50~\Mbps$.
Finally, we find that the cruising duration significantly limited the performance of UAV, as UAV may not have enough time slots to hover above Alice and Bob for data caching and delivery, respectively.

\begin{figure}[!t]
	\begin{center}
		\includegraphics[width=3.5 in]{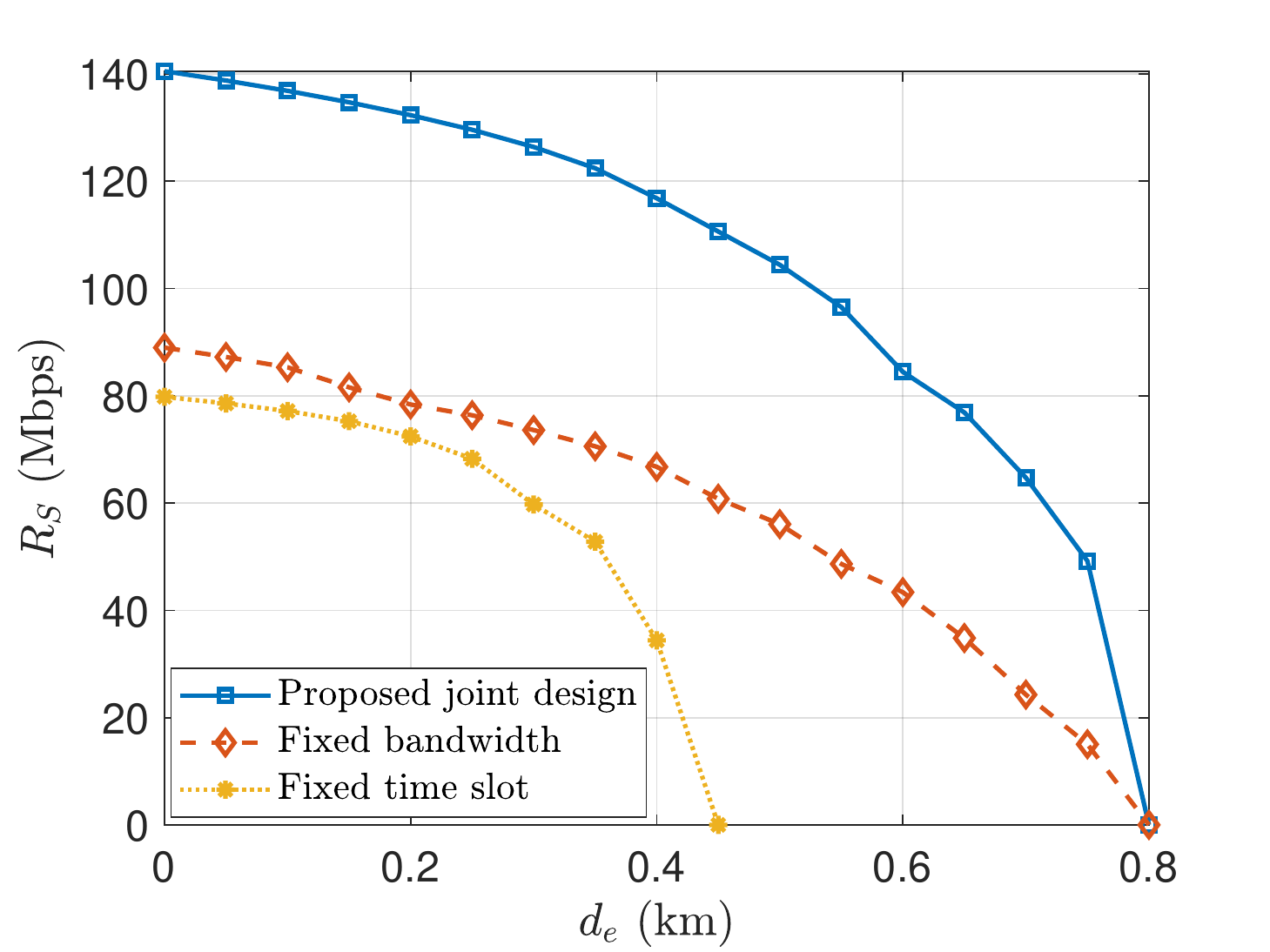} \vspace{-3ex}
		\caption{The maximal accumulated confidential data achieved at Bob versus uncertainty of Eve's location.}\label{fig:Compare}
	\end{center}
	\vspace{-5ex}
\end{figure}

In Fig. \ref{fig:Compare}, we compare the performance of the proposed joint design strategy with two strategies which serve as benchmarks.
In \textit{fixed bandwidth} strategy, the total bandwidth allocated to uplink and downlink transmission is fixed and equally divided, i.e., $b_{1,n}=0.5B$ and $b_{2,n}=0.5B, \forall n$.
In \textit{fixed time slot} strategy, the total time slots are equally allocated to Bob and Eve separately.
From this figure, we find that the proposed joint design strategy always outperforms the other two.
This is due to the fact that the proposed joint design exploits more degrees of freedom by adaptively adjusting the optimization variables.
For example, when the UAV approaches Alice for caching data, the proposed joint design tends to adaptively allocate more bandwidth resources to the uplink transmission.
Moreover, when secure communication can be guaranteed by optimizing the trajectory, the proposed design would allocate more time slots and power resources for transmitting confidential data as much as possible.
Furthermore, we find that the maximum accumulated confidential data rate achieved at Bob monotonically decreases with the increasing uncertainty of Eve's position.
The reasons are twofold.
With the increasing uncertainty of Eve's position, the UAV is forced to make a longer detour and reduce its transmit power for communicating with Bob, when it is close to the uncertain region leading to a smaller secrecy rate.
On the other hand, the resource allocation also tends to allocate more time slots and power to convey the messages to Eve for meeting its QoS requirement, which leads to a further decreasing performance achieved at Bob.

\section{Conclusion}

In this work, we have investigated the robust physical layer security against internal eavesdropping in a UAV-aided wireless communications system.
The optimization problem is formulated as a generalization of existing ones to improve the secrecy rate while guaranteeing a minimum required data rate for the internal eavesdropper by jointly designing UAV's trajectory and resource allocation.
To tackle the difficulties brought by the formulated non-convex NP-hard problem,
we have exploited the properties of the problem to simplify the formulation.
Then, a series of transformations and approximations have been proposed to facilitate the design of a computationally efficient suboptimal resource allocation algorithm.
Numerical results have demonstrated the advantages brought by the joint design for enhancing the physical layer security in communication  systems.

\renewcommand\refname{References}~\vspace{0mm}
\bibliographystyle{IEEEtran}
{\footnotesize\bibliography{IEEEabrv,Reference}}
\end{document}